\documentclass[preprintnumbers,prd,showpacs,floatfix,preprintnumbers,amsmath,amssymb,nofootinbib,superscriptaddress]{revtex4}
\usepackage{graphicx}
\usepackage{epsfig}
\usepackage{bm}
\usepackage{amsfonts}
\usepackage{color}
\usepackage{subfigure}
\usepackage{amsmath}
\usepackage{amssymb}
\usepackage{graphicx}
\usepackage{makecell}
\usepackage{float}

\usepackage{amsmath,amssymb}

\begin{document}

\title{Dark energy constraints from the 21~cm intensity mapping surveys with SKA1}

\author{Bikash R. Dinda}
\email{bikash@ctp-jamia.res.in} 
\affiliation{Centre for Theoretical Physics, Jamia Millia Islamia, New Delhi-110025, India}

\author{Anjan A Sen}
\email{aasen@jmi.ac.in} 
\affiliation{Centre for Theoretical Physics, Jamia Millia Islamia, New Delhi-110025, India}

\author{Tirthankar Roy Choudhury}
\email{tirth@ncra.tifr,res.in}
\affiliation{National Center For Radio Astrophysics, TIFR, Pst Bag 3, Ganeshkhind, Pune 411007, India}

\begin{abstract}
Understanding the nature of dark energy is one of the most outstanding problems in cosmology at present. In last twenty years, cosmological observations related to SNIa, Cosmic Microwave Background Radiation, Baryon Acoustic Oscillations etc, have put stringent constraints on the the dark energy evolution, still there is enough uncertainty in our knowledge about dark energy that demands new generation of cosmological observations. Post-reionization neutral hydrogen 21 cm intensity mapping surveys are one of the most promising future cosmological observations that have the potential to map the cosmological evolution from dark ages till present time with unprecedented accuracy and Square Kilometer Array (SKA) is one of the most sensitive instruments to measure the post-reionization 21 cm signal. In this work, we study the future dark energy constraints using post-reionization 21 cm intensity mapping power spectra with 
SKA1-mid specifications. We use three different parametrizations for dark energy equation of state (EoS) including the widely used CPL one. To generate simulated data, we use to two fiducial models: the concordance $\Lambda$CDM and the best fit CPL model for Planck+SNIa+BAO+HST. Our study shows that SKA1-mid alone has the potential to reach the  present accuracy for combined Planck+SNIa+BAO+HST to constrain the dark energy behaviour. Whether dark energy is  phantom or non-phantom or whether it exhibits phantom crossing, we may potentially address such questions with SKA1-mid. We also show that it is crucial to choose the correct parametrization for dark energy equation of state as some parametrizations are better than others to constrain the dark energy behaviour.  Specifically, as observed in this study, the widely used CPL parametrization may not give the best constraint for dark energy behaviour.   

\end{abstract}

\maketitle
\date{\today}

\section{Introduction}

During last two decades, we have witnessed some of the most accurate astrophysical and cosmological measurements which have established a standard model for cosmology with its parameters being measured with unprecedented accuracy \citep{riess, perlmutter, spergel, hinshaw, ade1,ade2, delubac, ata}. After the results from the Planck-2015 \citep{ade1, ade2}, it is remarkable that a concordance flat $\Lambda$CDM cosmology with only six parameters can describe our universe right from inflationary era till the present accelerating period including the periods when complicated processes like recombination and structure formation took place.

But few recent results from low redshift measurements in our universe showed inconsistencies with Planck-2015 results for flat $\Lambda$CDM model. Most prominent is the latest model independent measurement of $H_{0}$ by Riess et al. ( hereafter R16) \citep{R16} that has more than $3\sigma$ deviation from the Planck-2015 measurement for the same for a flat $\Lambda$CDM model. Same kind of inconsistency is also present in the $H_{0}$ measurement by strong lensing experiment like H0LiCow using time delay\citep{bonvin} and Planck-2015 measurement for $H_{0}$ for flat $\Lambda$CDM model. Do note that measurements of $H_{0}$ by R16 and H0LiCow are fully consistent.  Subsequently it was shown by Valentino et al. \citep{valentino} that this inconsistency between high redshift measurement of $H_{0}$ by Planck-2015 and low redshift measurements like R16 and H0LiCow, can be resolved by introducing a varying dark energy instead of a constant $\Lambda$.   Just recently Zhao et al. \citep{gongbo} have confirmed that a varying dark energy is preferred over cosmological constant at $3.5\sigma$ confidence level using a combination of low redshift and high redshift measurements ( also see Sahni et al. \citep{sahni} for model independent evidence for varying dark energy using only BAO data). To summarise, the fundamental question in cosmology at this moment that what drives the late time acceleration of the universe, is still far from being settled. 

The observational probes mentioned above, can be efficiently complemented by the upcoming neutral hydrogen (HI) 21~cm intensity mapping surveys \citep{bharadwaj1, wyithe}.  In particular, the Square Kilometer Array (SKA) will be one of the most sensitive instruments to detect the 21~cm signal from distant galaxies \citep{bull,ska1,ska2,ska3}. With a single set up, it is expected to provide the three dimensional information about the large scale structures in the universe. The observations of intensity mapping will essentially measure the large-scale fluctuations in the HI distribution at redshifts $z \lesssim 3$, which can, in principle, be analysed to understand the nature of dark energy that drives the late time acceleration in the universe. For our purpose, the relevant facility would be the SKA-Mid which is essentially a collection of radio dishes planned in South Africa.

In case one wants to study dark energy models that are different from the standard cosmological constant, the simplest way to model the time evolution of the dark energy is by using various kinds of scalar field model. These include, e.g., canonical scalar field \citep{quint}, non-minimal \citep{nonmin} and non-canonical scalar fields \citep{noncan}, phantom fields \citep{phantom}, galileons fields \citep{gal} to name a few. All these fields can also be subdivided into tracker/freezing and thawer classes depending on the nature of the initial conditions and the subsequent evolution \citep{lindcald}. But given a large class of scalar field models and its variants, it is always difficult to constrain individual models using observational data. Instead a suitable parametrisation of the broader behaviour of the dark energy in these models can be found and can be matched with the observations. The equation of state (EoS) for the dark energy is a reliable parameter that can describe the overall behaviour of any dark energy model. Many parametrisations for the equation of state for the dark energy haven been proposed \citep{cpl,gcg,ba,jbp,7cpl}. The parametrisation by Chevallier, Polarski and Linder ( hereafter CPL) \citep{cpl} is one such parametrisation that has been used universally to fit dark energy with observational data.

The main aim of this work is to use the CPL parametrisation and two of its variants, namely the 7CPL \citep{7cpl}and the BA \citep{ba}, together with the concordance $\Lambda$CDM model, to investigate the future prospects SKA HI intensity mapping survey to constrain the behaviour of the dark energy. Forecasting the constraints on late time cosmology using 21cm intensity mapping experiments has been previously done by Bull et al. \citep{Bull} using $\Lambda$CDM and dark energy with CPL parametrization only. Prospects of probing quintessence 2D angular power spectra for intensity mapping  has also been studied by Hussain et al. \citep{PS_21_6}. The specifications for SKA configurations have been revised in recent past. Here we use the revised specifications for SKA1-mid for our study. We also consider different parametrizations for dark energy equation of state (EoS) to see which parametrization gives the best constraints for dark energy evolution.   

The structure of the paper is as follows: in Section II, we describe the background cosmology and different parametrizations for dark energy equation of state; Section III deals with the linear density perturbations and the linear matter power spectra for different dark energy models considered; Section IV deals with the observed 3d 21 cm intensity mapping power spectra; in Section V, we describe the method of calculating the system noise and sample variance and the specifications for SKA1-mid to calculate such quantities; in Section VI, we describe the results obtained in this work and finally we discuss our conclusions in Section VII.

\section{Background evolution with dark energy parametrizations}

The energy components relevant for studying the evolution of the Universe post-recombination are the non-relativistic matter (which includes both cold dark matter and baryons) and the dark energy. Using a spatially flat FRW metric for the background Universe, the background evolution of the Hubble parameter $ H $ is given by
\begin{equation}
H(z) = H_{0} \sqrt{\Omega_{m}^{(0)} (1+z)^{3} + (1-\Omega_{m}^{(0)}) \exp \Big{[} 3 \int_{0}^{z} \frac{1+w(z')}{1+z'} dz' \Big{]}},
\label{eq:hubble}
\end{equation}
where $ z $ is the redshift, $ \Omega_{m}^{(0)} $ and $ H_{0} $ are the present day matter energy density parameter and Hubble parameter respectively and $ w(z)$ is the equation of state (EoS) of the dark energy. For $w(z)$, we consider various parametrizations as following: 
\begin{eqnarray}
w(z) &=& w_{0}^{CPL}+w_{a}^{CPL} \frac{z}{1+z} \hspace{0.5 cm} \text{(CPL)} \nonumber\\
w(z) &=& w_{0}^{BA}+w_{a}^{BA} \frac{z(1+z)}{1+z^{2}} \hspace{0.5 cm} \text{(BA)} \nonumber\\
w(z) &=& w_{0}^{7CPL}+w_{a}^{7CPL} \Big{(} \frac{z}{1+z} \Big{)}^{7} \hspace{0.5 cm} \text{(7CPL)},
\label{eq:eos}
\end{eqnarray}
where $ w_{0} $ and $ w_{a} $ are two model parameters for each parametrization. The parameter $ w_{0} $ represents the present day ($z = 0$) value of the dark energy e.o.s and $ w_{a} $ represents its redshift evolution at present ($z=0$). Among these three parametrizations, the CPL and BA parametrizations can mimic thawing class of dark energy models by proper adjustment of the parametrs $ w_{0} $ and $ w_{a} $ whereas the 7CPL parametrization can mimic the tracker class of dark energy models better.

\section{Evolution of the linear perturbation}

In this section, we study the evolution of the matter density perturbations in the dark energy models under the Newtonian approximation. The approximation is valid as long as we are interested in scales smaller than the horizon. In this regime, we can safely ignore the effect of dark energy perturbations as they are only affect the clustering at horizon scales and beyond. The evolution of the matter density contrast in the linear approximation is given by
\begin{equation}
\ddot{\delta}_{m} + 2 H \dot{\delta}_{m} - 4 \pi G \bar{\rho}_{m} \delta_{m} = 0,
\label{eq:delm}
\end{equation}
where $ \bar{\rho}_{m} $ is the background matter energy density and $ G $ is the Newtonian gravitational constant. The matter density contrast is defined as $ \delta_{m} = \rho_{m} / \bar{\rho}_{m} - 1 $ where $ \rho_{m} $ is the total matter energy density.  The growing solution of the above equation defines the growth function of the matter perturbations $ D_{m} $. Hence the growth function evolution equation w.r.t $ N = \ln a $ can be written as
\begin{equation}
\dfrac{d^{2} D_{m}}{d N^{2}} + \frac{1}{2} \Big{(} 1 - 3 w \Omega_{q}(z) \Big{)} \dfrac{d D_{m}}{d N} - \frac{3}{2} \Omega_{m}(z) D_{m} = 0,
\label{eq:Dm}
\end{equation}
where $ \Omega_{m} (z)= 1 -\Omega_{q}(z) $ is the matter energy density parameter. To solve the above equation, we assume $ D_{m} \propto a $ at early matter dominated era with $ a $ being the scale factor. Therefore we choose our initial conditions as $ D_{m}^{i} = a_{i} = d D_{m} / d N \big{|}_{i} $ with $a_{i} = 0.001$.

%%%%%%%%%%%%%%%%%%%%%%%%%%%%%%%%%%
\begin{figure*}[tbp]
\begin{center}
\includegraphics[width=.45\textwidth]{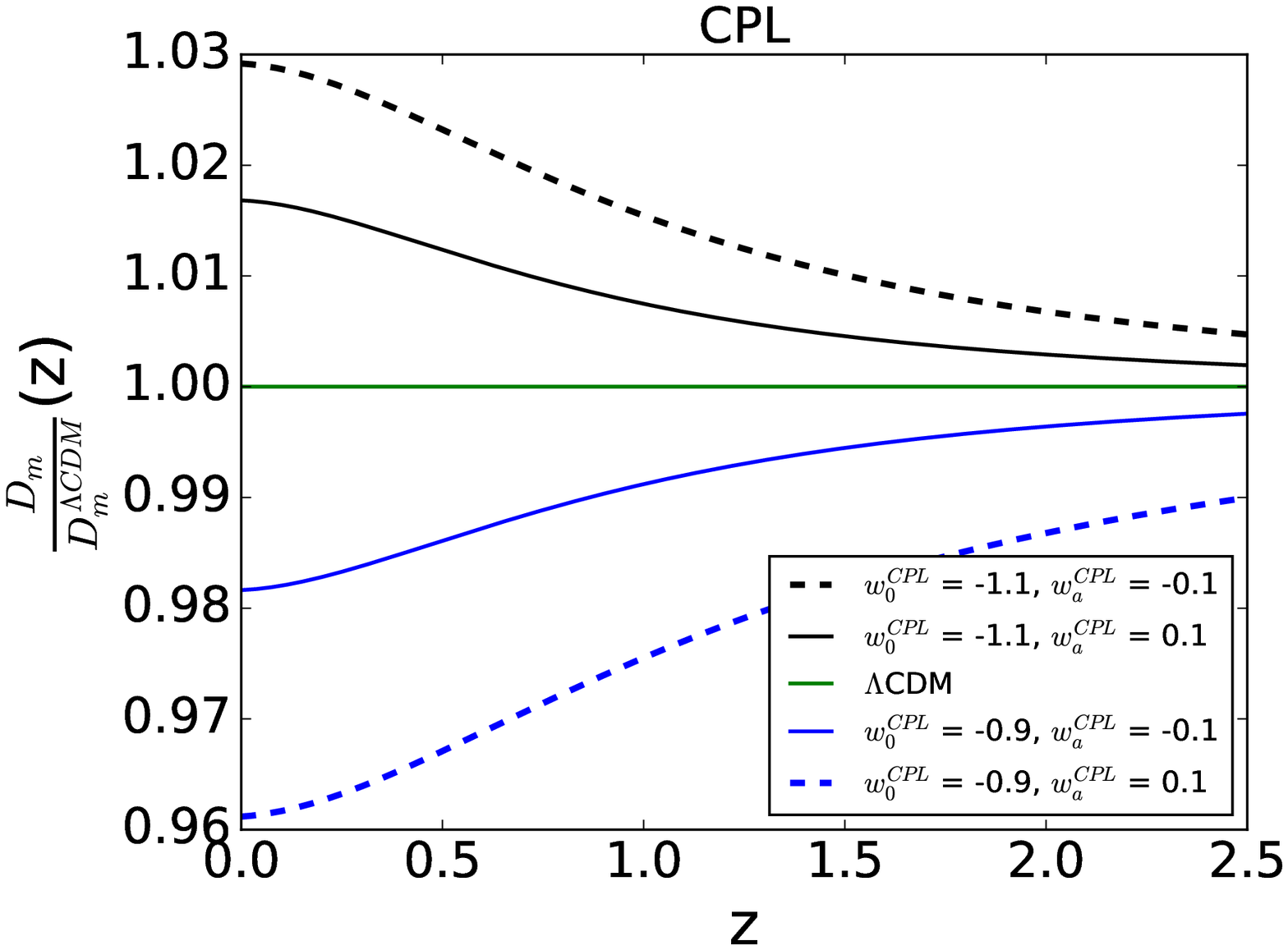}
\includegraphics[width=.45\textwidth]{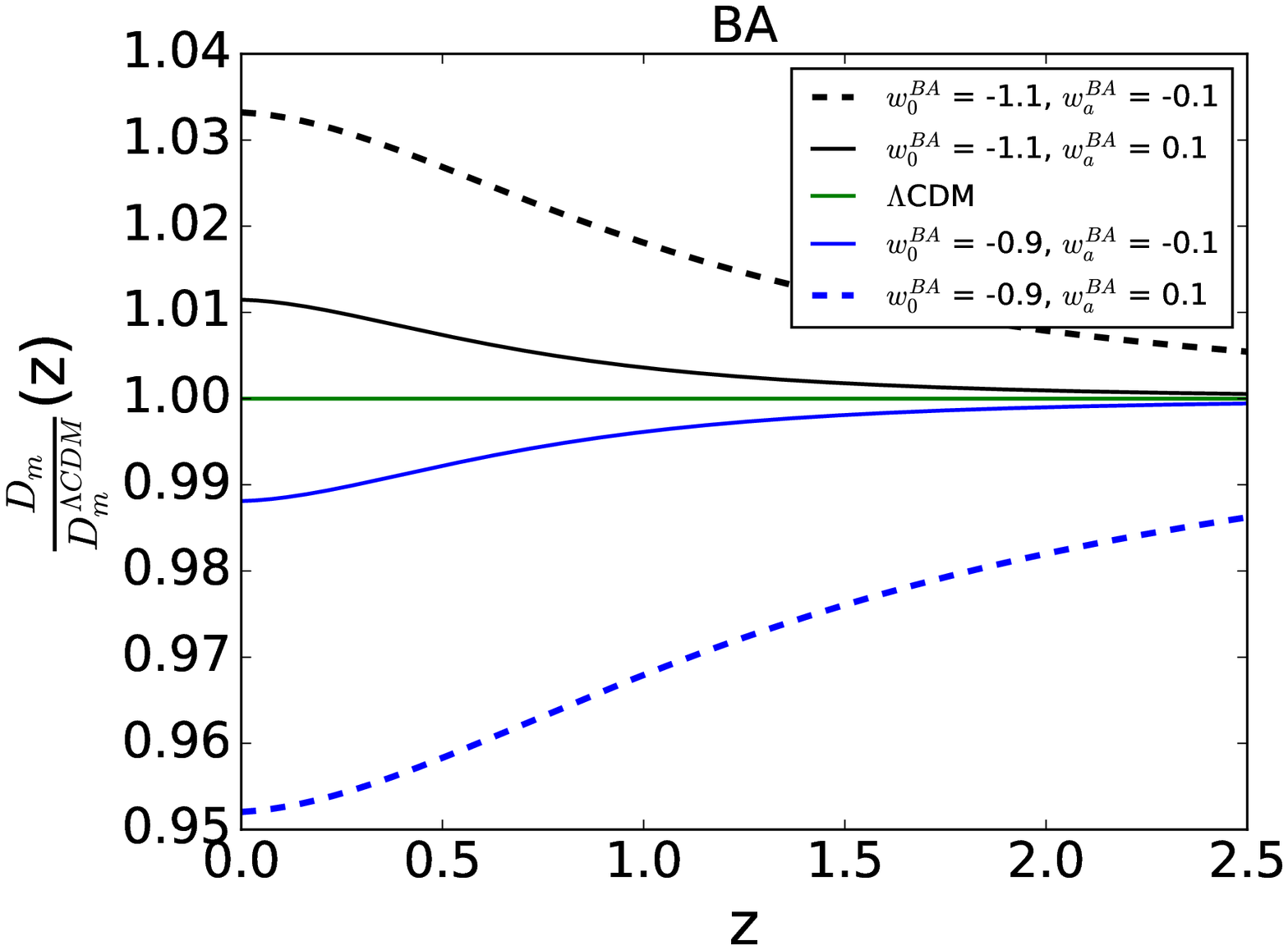}\\
\includegraphics[width=.45\textwidth]{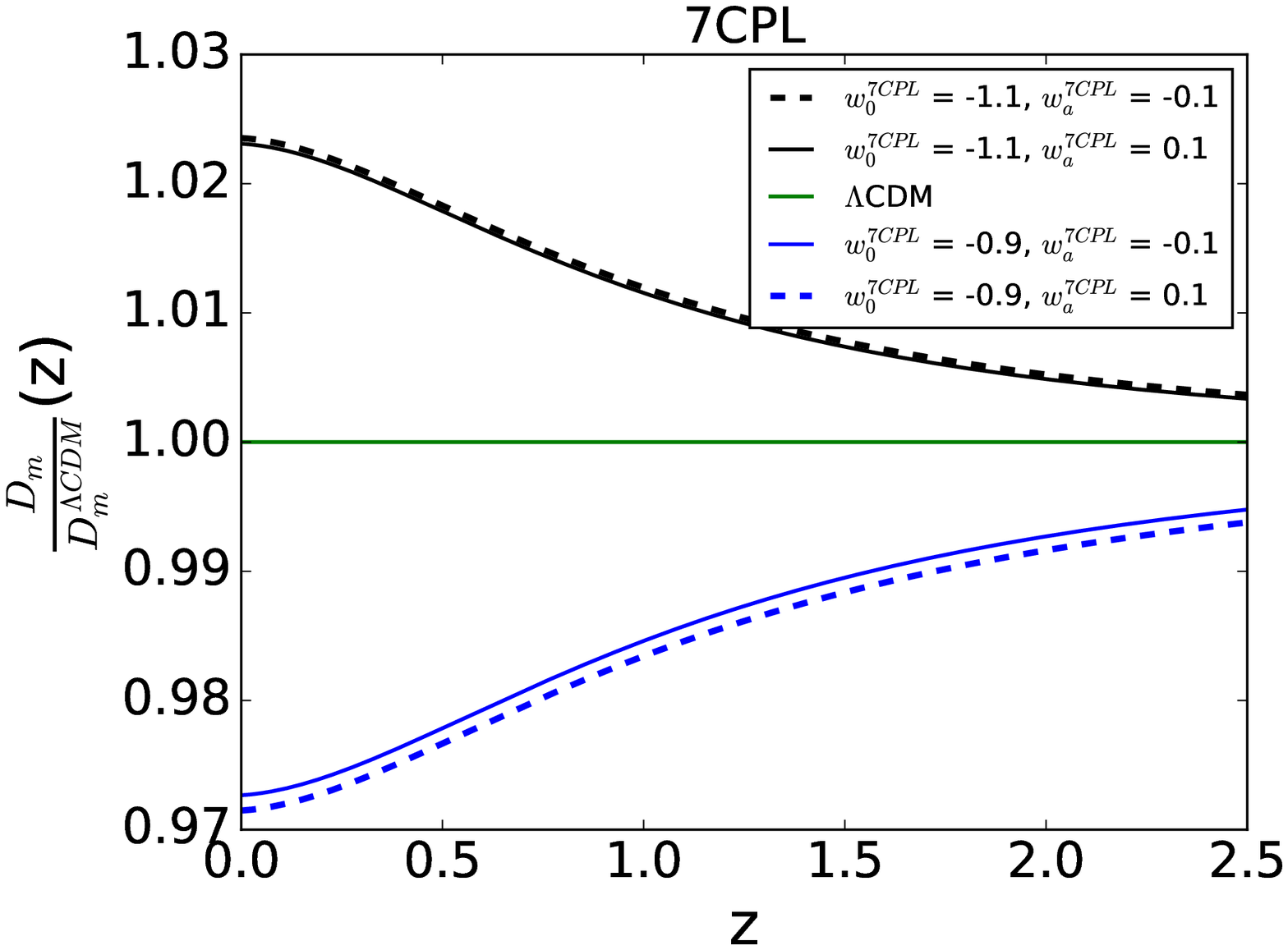}
\caption{\label{fig:growth} Deviation in growing mode for linear growth function from $ \Lambda $CDM for three different parametrizations.}
\end{center}
\end{figure*}
%%%%%%%%%%%%%%%%%%%%%%%%%%%%%%%%%%

In Figure~\ref{fig:growth}, we plot the deviation in the linear growth function for four different parametrizations from that in the $ \Lambda $CDM model. At higher redshifts, the clustering in all the models approach to $\Lambda$CDM behaviour as expected. For smaller redshifts, clustering in matter is larger for phantom models ( $w_{0} < -1$) compared to $\Lambda$CDM whereas for non-phantom models, it is smaller compared to $\Lambda$CDM. Interestingly for 7CPL model, there is no dependence of the parameter $w_{a}$ on matter clustering. Therefore one does not expect to constrain the parameter $w_{a}$ for 7CPL model using any observation related to matter clustering.

The linear matter power spectrum $ P_{m} $ is defined as
\begin{equation}
P_{m}(k,z) = A_{H} k^{n_{s}} T^{2} (k) \frac{D_{m}^{2}(z)}{D_{m}^{2}(z=0)},
\label{eq:Pm}
\end{equation}
where $ k $ is the amplitude of the wave vector $ \vec{k} $ and $ n_{s} $ is the scalar spectral index corresponding to the primordial curvature perturbation. $ T(k) $ is the transfer function. For our calculations, we consider the $T(k)$ given by Eisenstein and Hu \citep{eisenhu}. $ A_{H} $ is the normalisation constant which is determined by the usual $ \sigma_{8} $ normalisation. As before, we fix $ \Omega_{m}^{(0)} = 0.308 $ and $ h = 0.69 $ for the background evolution. In addition, we also fix the values of other cosmological parameters to $ \Omega_{b}^{(0)} = 0.05 $, $ n_{s} = 0.968 $ and $ \sigma_{8} = 0.8 $. These are within $1\sigma$ bound by Planck-2015 for $\Lambda$CDM model \citep{ade1}.

\section{Observed 21 cm power spectrum}

According to our current understanding, the post reionization ($z \lesssim 6$) 21~cm power spectrum arises from the neutral hydrogen within galaxies. While the knowledge about the distribution of HI within galaxies is limited, fortunately, such details are less important for modelling the 21~cm fluctuations at cosmological scales. At sufficiently large scales, one can assume that the HI fluctuations trace the underlying dark matter density field with an appropriate bias factor included.

The observable in 21~cm experiments is the excess brightness temperature. The power spectrum $ P_{21} $ of this excess brightness temperature field is given by \citep{PS_21_1,PS_21_2,PS_21_3,PS_21_5,PS_21_6,PS_21_7,PS_21_8,PS_21_10}

\begin{equation}
P_{21} (k,z,\mu) = C_{T}^{2} (1+\beta_{T} \mu^{2})^{2} P_{m}(k,z),
\label{eq:P212D}
\end{equation}

\noindent
where $ P_{m}(k,z) $ is the matter power spectrum defined in eq. \eqref{eq:Pm}, $ \mu $ is the cosine of the angle between line of sight ($ \vec{n} $) and the wave vector ($ \vec{k} $), i.e., 
\begin{equation}
\mu = \hat{n}.\hat{k} = \frac{k_{||}}{k},
\end{equation}
where $k_{||}$ is the component of $\vec{k}$ along the line of sight. The parameter $ \beta_{T} = f_{m} / b_{T} $ is the redshift distortion factor, with $ b_{T} $ being the linear bias and $ f_{m} $ being the linear matter growth rate. The quantity $f_m$ is defined as $ d \ln D_{m} / d N $. $C_{T} $ is the mean 21 cm excess brightness temperature, given by \citep{PS_21_1,PS_21_2}
\begin{equation}
C_{T} (z) = b_{T} \bar{x}_{HI} \bar{T} (z).
\label{eq:CT}
\end{equation}
In the above expression, $ \bar{x}_{HI} $ is the neutral hydrogen fraction and $ \bar{T} (z) $ is given by \citep{PS_21_1,PS_21_2}
\begin{equation}
\bar{T} (z) = 4.0 mK (1+z)^{2} \Big{(} \frac{\Omega_{b0} h^{2}}{0.02} \Big{)} \Big{(} \frac{0.7}{h} \Big{)} \frac{H_{0}}{H(z)}.
\label{eq:Tbar}
\end{equation}

In actual radio interferometric observations of the 21~cm signal, the observables are measured in terms of baselines $\vec{U}$ and the frequency $\nu$. To convert these two variables to $k_{\perp}$ and $k_{||}$ (or equivalently to $k$ and $\mu$), one needs to assume a ``fiducial'' cosmological model. While comparing the observations with predictions from different cosmological models, it is important to account for the fact that the fiducial model will, in general, be different from the cosmological model under consideration. In such a case, the observed 21 cm power spectrum will be given by \citep{Bull,P21_3D_2,P21_3D_5}
\begin{equation}
P_{21}(k,z,\mu) = \frac{1}{\alpha_{||} \alpha_{\perp}^{2}} C_{T}^{2} \Big{[} 1+\beta_{T} \frac{\mu^{2} / F^{2}}{1+(F^{-2}-1) \mu^{2}} \Big{]}^{2} P_{m} \Big{(} \frac{k}{\alpha_{\perp}} \sqrt{1+(F^{-2}-1) \mu^{2}} \Big{)},
\label{eq:P213D}
\end{equation}
where $ \alpha_{||} = H_{\rm fd}(z) / H(z) $, $ \alpha_{\perp} = r(z) / r_{\rm fd}(z) $ and $ F = \alpha_{||} / \alpha_{\perp} $. The subscript 'fd' represents quantities for the fiducial model. Note that the above expression is the equivalent of equation (17) of \citep{Bull}, keeping in mind that the definitions of $\alpha_{||}$ and $\alpha_{\perp}$ are different in our analysis.

The angle averaged 21 cm power power spectrum (usually known as the monopole) is given by
\begin{eqnarray}
P_{21}(k,z) = \int_{0}^{1} d\mu \hspace{0.2 cm} P_{21}(k,z,\mu).
\label{eq:P213Davg}
\end{eqnarray}
Note that for simplicity, we use the same notation $ P_{21} $ in the three equations \eqref{eq:P212D}, \eqref{eq:P213D} and \eqref{eq:P213Davg}. In all the subsequent figures, $ P_{21} $ implies the one which is used in eq \eqref{eq:P213Davg}. 

%%%%%%%%%%%%%%%%%%%%%%%%%%%%%%%%%%
\begin{figure}[tbp]
\centering
\includegraphics[width=.45\textwidth]{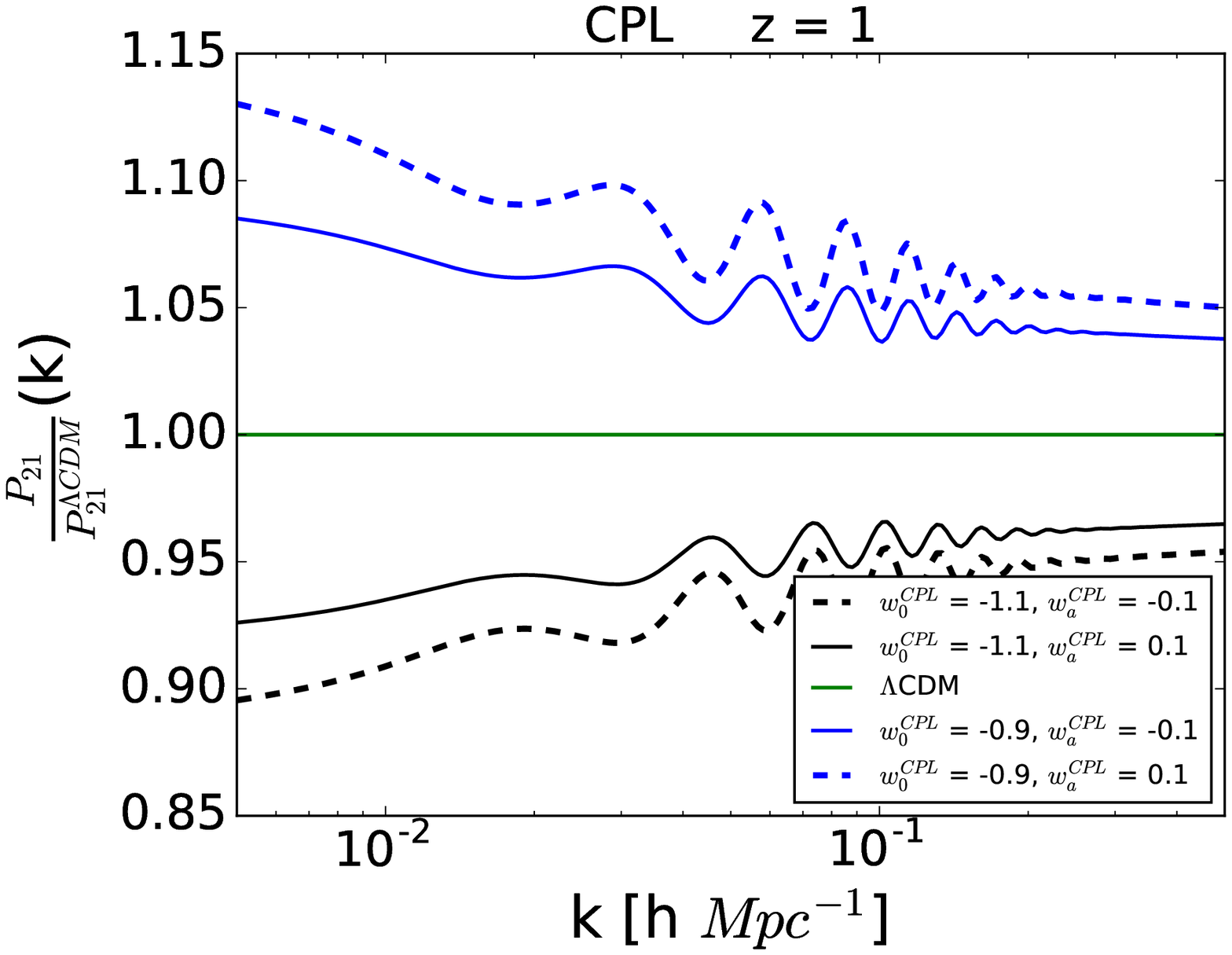}
\includegraphics[width=.45\textwidth]{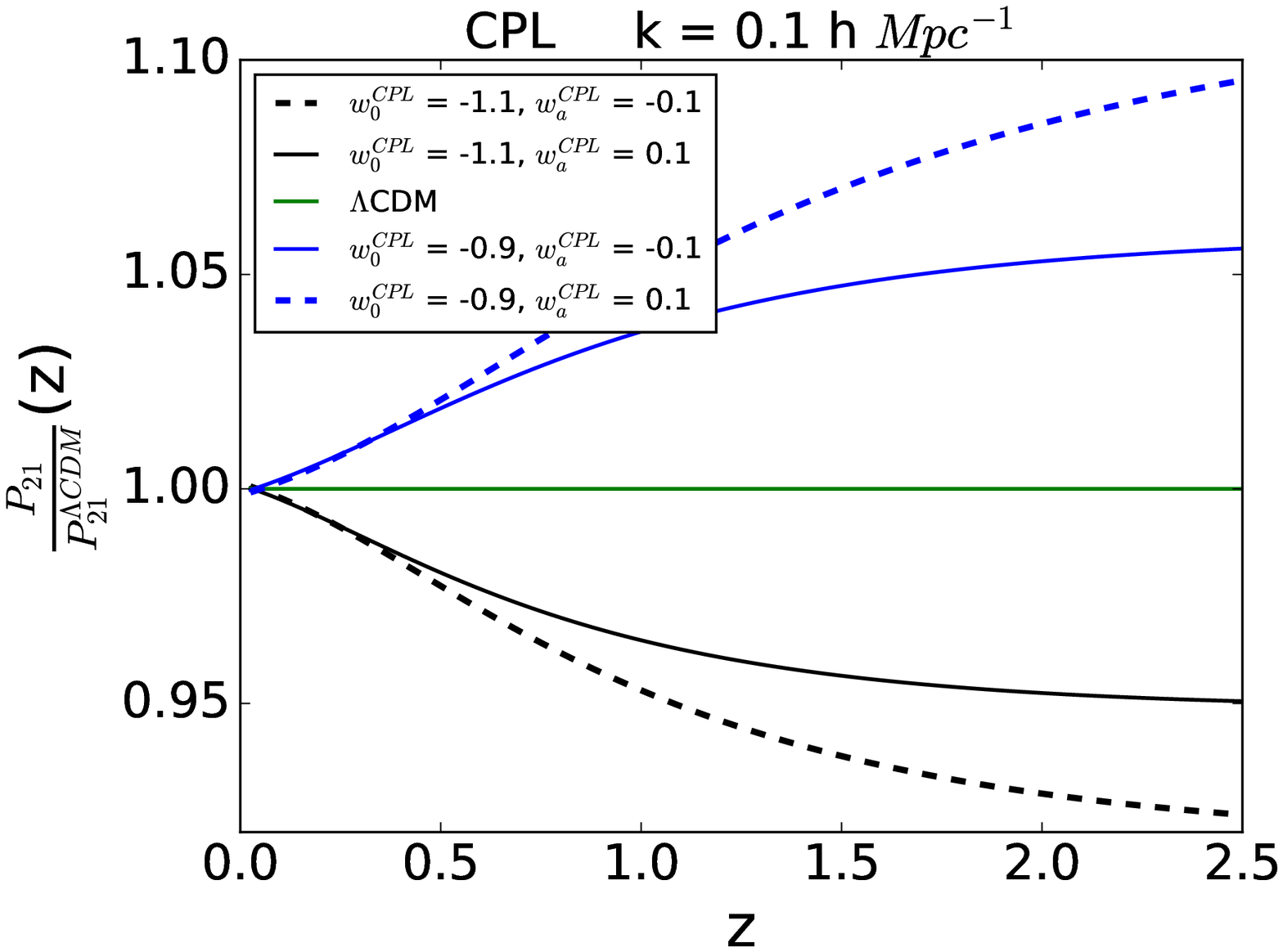}\\
\includegraphics[width=.45\textwidth]{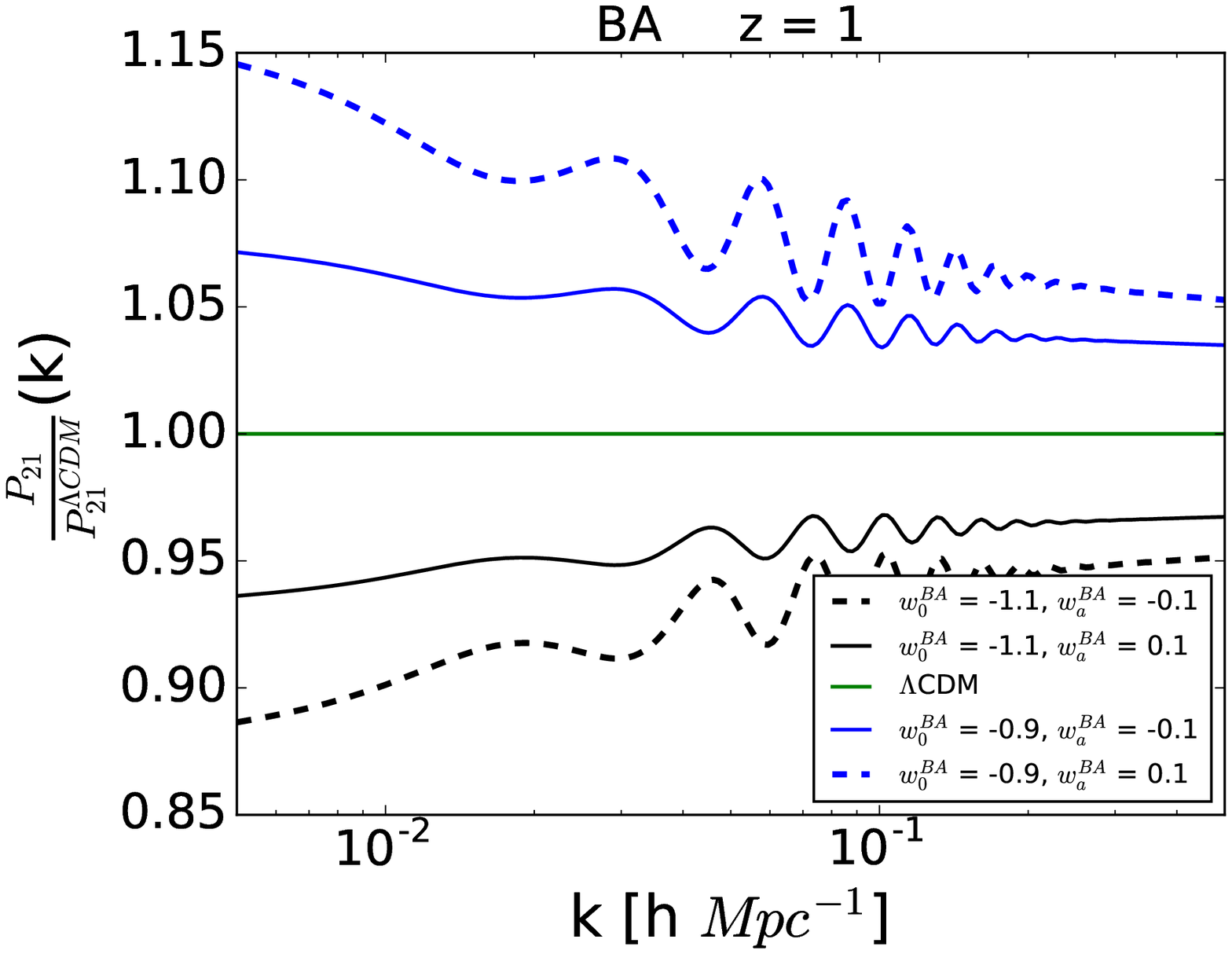}
\includegraphics[width=.45\textwidth]{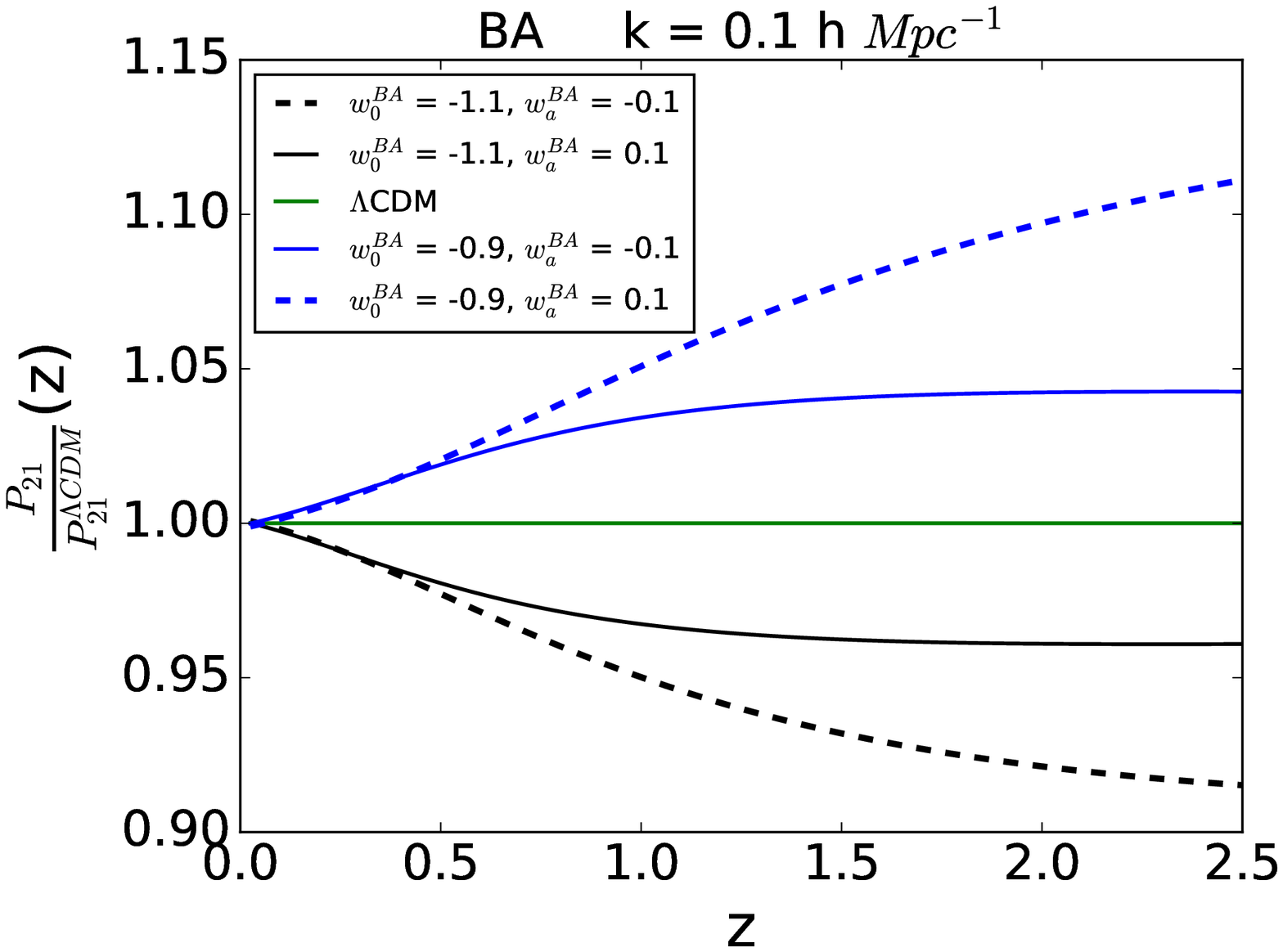}\\
\includegraphics[width=.45\textwidth]{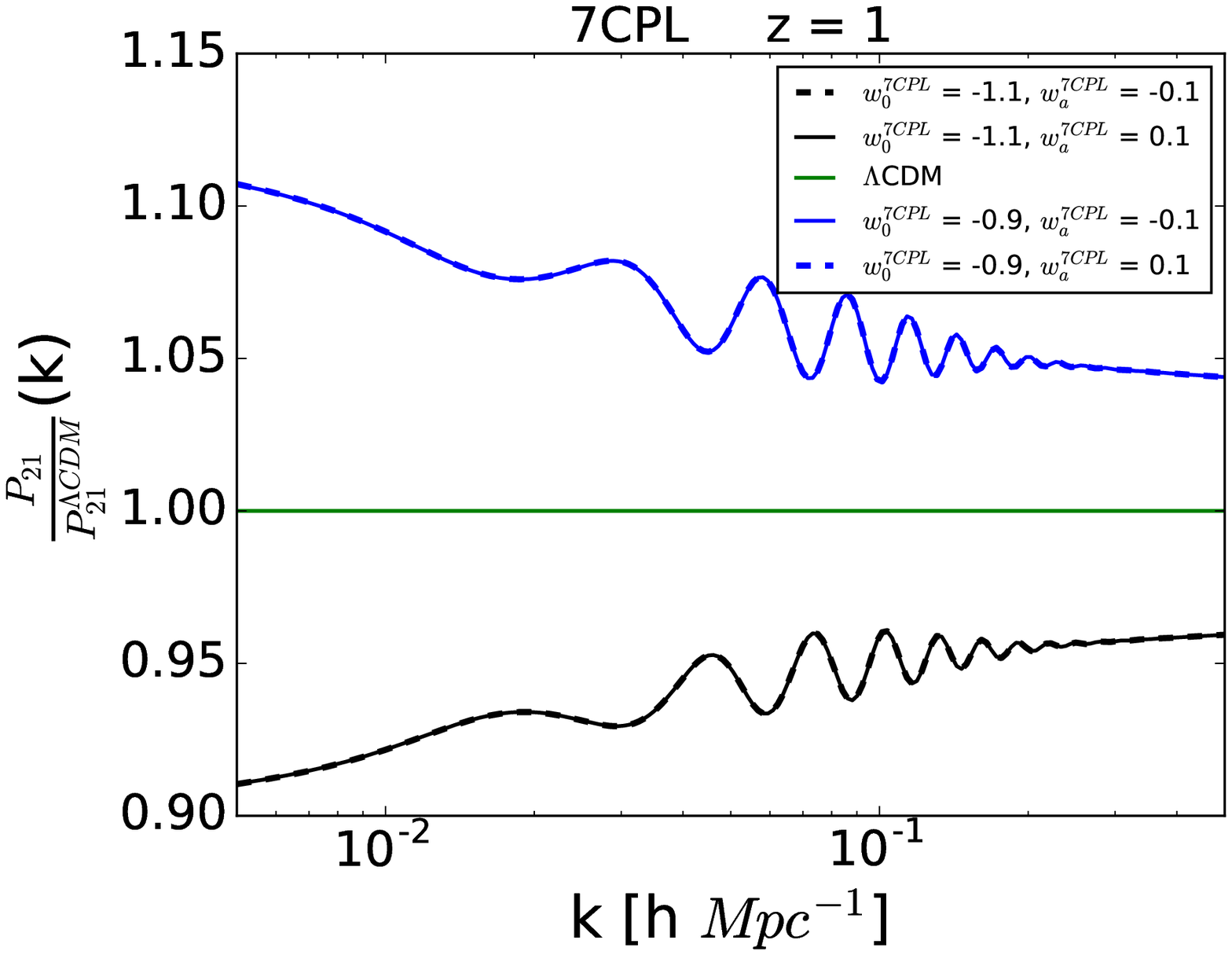}
\includegraphics[width=.45\textwidth]{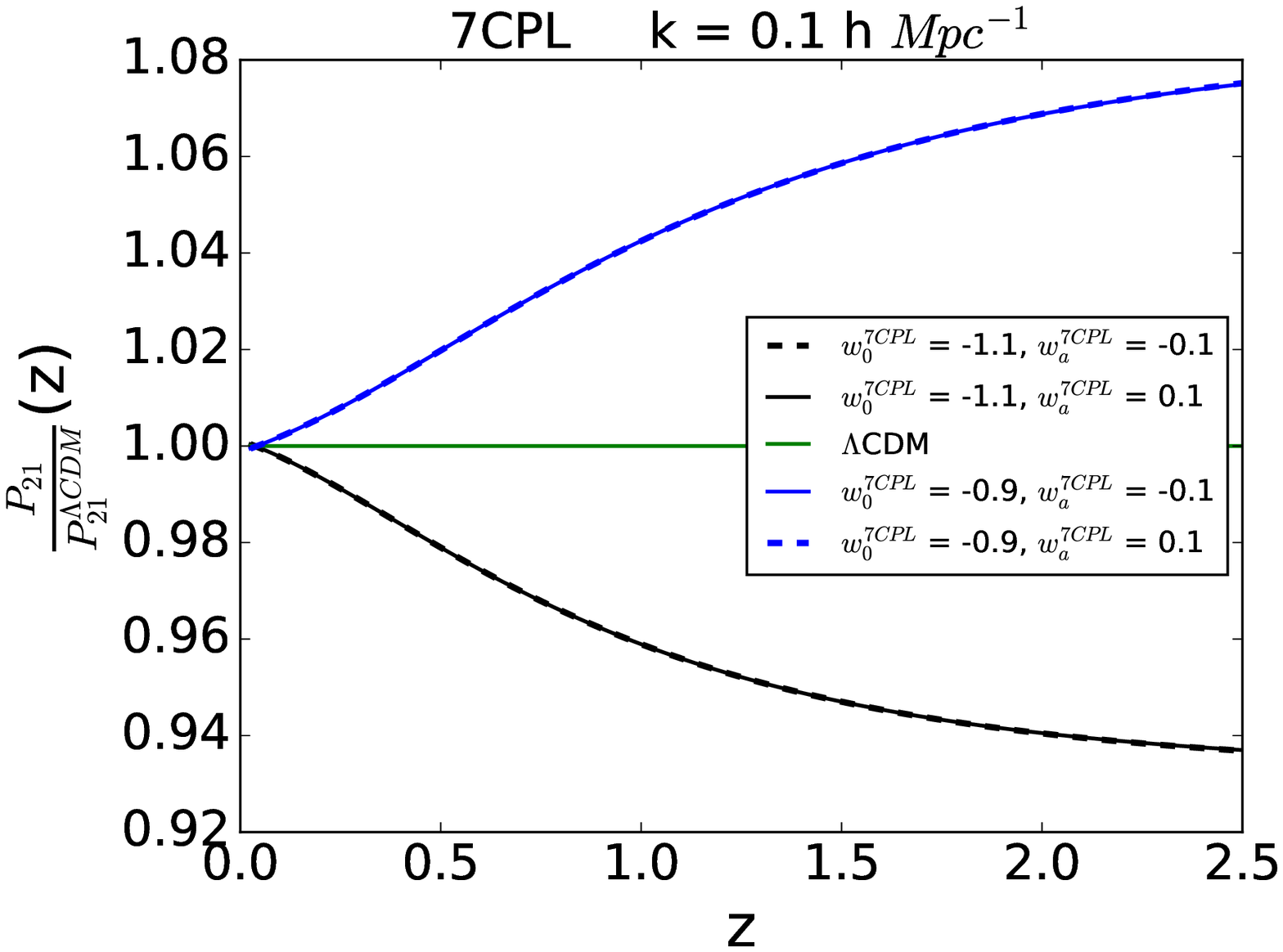}
\caption{\label{fig:ps21D3Lcdm} Deviation in the 21 cm power spectrum from $ \Lambda $CDM model for different parametrizations.}
\end{figure}
%%%%%%%%%%%%%%%%%%%%%%%%%%%%%%%%%%

In Figure~\ref{fig:ps21D3Lcdm}, we show the deviation in the observed 21 cm power spectrum from $ \Lambda $CDM model for the four different parametrizations (see equation (2)). The fiducial model has been assumed to be the $ \Lambda $CDM. We can see from the figure that the deviations are dependent on the scale considered unlike for matter power spectrum (see equation (5)) where deviation in any model from $\Lambda$CDM depends only on the corresponding deviation in growth factor which is scale independent. The scale-dependence in the observed 21 cm power spectrum arises due to the fact that the parallel and the perpendicular components of the wave vector $ \vec{k} $ are scaled by different factors $ \alpha_{||} $ and $ \alpha_{\perp} $ respectively. This in turn is related to the fact that radio interferometric observations have different responses in the sky plane and along the frequency direction. We also see that the deviations increase with increasing redshift and decrease for smaller scales. As expected, the deviations do not depend on the $ w_{a}^{7CPL} $ parameter for a fixed value of $ w_{0}^{7CPL} $ for the 7CPL parametrization. For all the three parametrizations, there is always more power than $\Lambda$CDM for non-phantom cases whereas the power is less than the $\Lambda$CDM for phantom cases.

\section{System Noise and Sample Variance}

The errors on the 21~cm power spectrum measurements arise mainly from two sources, namely the system noise arising from the telescope and the cosmic variance arising because of finite sampling of modes. In addition, errors can arise from the residual astrophysical foregrounds. In this work, we ignore the effect of foregrounds and assume them to be completely removed. There will also be systematic errors because of the ionosphere and man-made interference at radio frequencies. These sources of uncertainties too are ignored in the present analysis and we assume a rather ideal instrument than what is expected in reality.

\subsection*{System Noise}

The system noise depends on the configuration of the telescope and the observational settings employed or detecting the signal. In this work, we follow the approach of \citep{noise_err_1} to estimate the noise.

Let us, for simplicity, assume a circularly symmetric antenna distribution denoted by $ \rho_{ant} (l) $ where $ l $ is the distance from the centre of the area where antennae are located. This approximation is reasonable for antennae near the core of the telescope for facilities like the SKA1. From this antenna distribution one gets the 2-D baseline distribution given by \citep{noise_err_1}
\begin{equation}
\rho_{2D} (U,\nu) = B(\nu) \int_{0}^{\infty} 2 \pi l dl \hspace{0.1 cm} \rho_{ant} (l) \int_{0}^{2 \pi} d\phi \hspace{0.1 cm} \rho_{ant} (|\vec{l}-\lambda \vec{U}|),
\label{eq:rho2d}
\end{equation}
\noindent
where the circular symmetry implies that $ \rho_{2D} $ is a function only of $ |\vec{U}| = U $. The baseline vector $ \vec{U} $ related to the  wave vector is given by $ \vec{U} = \frac{\vec{k}_{\perp} r_{\nu}}{2 \pi} $ where $ r_{\nu} $ be the comoving distance at a redshift $ z $ corresponding to the observed frequency $ \nu $ of the 21 cm signal and $ \vec{k}_{\perp} $ is the component of the wave vector $ \vec{k} $ in the perpendicular direction of the line of sight (i.e., in the sky plane). The magnitude of $ \vec{k}_{\perp} $ is given by $ k_{\perp} = k \sin\theta $ whereas the parallel component is given by $ k_{||} = k \cos\theta $. The  $ B(\nu) $ is the normalised constant which is fixed by the normalisation given by
\begin{equation}
\int_{0}^{\infty} U dU \int_{0}^{\pi} d\phi \hspace{0.1 cm} \rho_{2D} (U,\nu) = 1.
\label{eq:rho2dnorm}
\end{equation}
Note that the normalisation constant $ B(\nu) $ has different values at different redshifts corresponding to the frequency $ \nu $. The system noise error ($ 1 \sigma $) in the power spectrum for a single mode having wave vector $ \vec{k} $ is given by
\begin{eqnarray}
\delta P_{N}^{1-mode} (U,\nu) &=& \frac{T_{sys}^{2}}{B t_{0}} \Big{(} \frac{\lambda^{2}}{A_{e}} \Big{)}^{2} \frac{r_{\nu}^{2} L}{n_{b} (U,\nu)} \nonumber\\
&=& \frac{T_{sys}^{2}}{B t_{0}} \Big{(} \frac{\lambda^{2}}{A_{e}} \Big{)}^{2} \frac{2 r_{\nu}^{2} L}{N_{t} (N_{t}-1) \rho_{2D} (U,\nu)},
\label{eq:pnsm}
\end{eqnarray}
where $ T_{sys} $ is the system temperature of the instrument, $ B $ is the total frequency bandwidth, $ t_{0} $ is the observation time, $ L $ is the comoving length corresponding to the bandwidth $ B $, and $ \lambda $ is the observed wavelength corresponding to the observed frequency $ \nu $ of the 21 cm signal. The quantity $ A_{e} $ is the effective collecting area of an individual antenna which can be written as $ A_{e} = \epsilon A $, where $ \epsilon $ is the efficiency of an antenna and $ A $ is the physical collecting area of an individual antenna. The relationship between $ n_{b} $ and $ \rho_{2 D} $ is given by
\begin{equation}
n_{b} (U,\nu) = \frac{N_{t} (N_{t}-1)}{2} \rho_{2D} (U,\nu),
\label{eq:rho2dnorm}
\end{equation}
where $ N_{t} $ is the total number of antennae. Now the average error in the power spectrum for all the modes which lie in the range $ k $ to $ k + dk $ and $ \theta $ to $ \theta + d \theta $ is given by
\begin{equation}
\delta P_{N} (U,k,\theta,\nu) = \frac{T_{sys}^{2}}{B t_{0}} \Big{(} \frac{\lambda^{2}}{A_{e}} \Big{)}^{2} \frac{2 r_{\nu}^{2} L}{N_{t} (N_{t}-1) \rho_{2D} (U,\nu)} \frac{1}{\sqrt{N_{m}(k,\theta)}},
\label{eq:pn1}
\end{equation}
where $ \theta $ is the angle between $ \vec{k} $ and the line of sight direction and the total no of independent modes in the range $ k $ to $ k + dk $ and $ \theta $ to $ \theta + d \theta $ is given by

\begin{equation}
N_{m}(k,\theta) = \frac{2 \pi k^{2} dk \sin\theta d\theta}{V_{1-mode}},
\label{eq:nm}
\end{equation}

\noindent
where

\begin{equation}
V_{1-mode} = \frac{(2 \pi)^{3} A}{r_{\nu}^{2} L \lambda^{2}},
\label{eq:V1md}
\end{equation}

\noindent
is the volume occupied by a single independent mode in k-space. Now we define a quantity $ \rho_{3D}(k,\nu) $ which is related to the 2-D normalised baseline distribuation given by

\begin{eqnarray}
\rho_{3D}(k,\nu) &=& \Big{[} \int_{0}^{\frac{\pi}{2}} d\theta \sin\theta \rho_{2D}^{2} \big{(} \frac{r_{\nu} k}{2 \pi} \sin\theta,\nu \big{)} \Big{]}^{\frac{1}{2}} \nonumber\\
&=& \Big{[} \int_{0}^{1} d\mu \hspace{0.2 cm} \rho_{2D}^{2} \big{(} \frac{r_{\nu} k}{2 \pi} \sqrt{1-\mu^{2}},\nu \big{)} \Big{]}^{\frac{1}{2}},
\label{eq:rho3d}
\end{eqnarray}

\noindent
where $ \mu = \cos\theta $. Using above equation we finally get the system noise error in the 21 cm power spectrum in the continuum limit given by \citep{noise_err_1,noise_err_2,noise_err_3,noise_err_4}

\begin{equation}
\delta P_{N} (k,\nu) = \frac{T_{sys}^{2}}{B t_{0}} \Big{(} \frac{\lambda^{2}}{A_{e}} \Big{)}^{2} \frac{2 r_{\nu}^{2} L}{N_{t} (N_{t}-1) \rho_{3D} (k,\nu)} \frac{1}{\sqrt{N_{k}(k)}}.
\label{eq:pnfinal}
\end{equation}

\noindent
where $ N_{k} $ is the total no of independent observable modes in the spherical shell between $ k $ to $ k + dk $ is given by

\begin{equation}
N_{k}(k) = \frac{2 \pi k^{2} dk}{V_{1-mode}}.
\label{eq:nk}
\end{equation}

\subsection*{Sample Variance}

Since only a finite number of modes will be probed in each $k$-bin, this will lead to a statistical error known as the sample (or cosmic) variance. In this work, we assume this error to be Poissonian, i.e., the error decreases with the number of modes as $N_m^{-1/2}$. Hence, the sample variance in 21 cm power spectrum is given by \cite{noise_err_1,noise_err_2,noise_err_3,noise_err_4}

\begin{equation}
\delta P_{SV} (k,\nu) = \Big{[} \sum_{\theta} \frac{N_{m}(k,\theta)}{P_{21}^{2}(k,\theta)} \Big{]}^{- \frac{1}{2}}.
\label{eq:sv}
\end{equation}
This error is usually important at very large scales where the number of modes probed are relatively smaller.

Considering only the system noise and the sample variance the total noise is thus given by 

\begin{equation}
\delta P_{tot} (k,\nu) = \delta P_{N} (k,\nu) + \delta P_{SV} (k,\nu).
\label{eq:totalN}
\end{equation}
We use the above expression to calculate the expected errors in the upcoming low-frequency telescopes.

\subsection*{Specifications for the SKA1-MID}

In this work, we will focus on constraining the dark energy models using a telescope like the SKA1-MID. Although the SKA project is going through various levels of descoping, hence the telescope properties used in this paper may still get modified. We assume that the SKA1-MID will have roughly a total 200 number of antannae (which includes 60 MeerKAT dishes). The diameter of each antenna is taken to be 15 meter. We take typical antenna efficiency to be $ \epsilon = 0.7 $. For the calculations of the system noise, we assume system temperature will be 40 K constant over the redshift range $ z \lesssim 2.5 $.

For the SKA1-MID we have taken the baseline distribution, $\rho_{2D} (U)$ given in the document \url{http://www.skatelescope.org/wp-content/uploads/2012/07/SKA-TEL-SKO-DD-001-1_BaselineDesign1.pdf} (see violet line of the figure (10) in this document).

The redshifts of our interest where the main constraints are expected will be around $z \sim 0.5 - 2.5$. This corresponds to a frequency range $400 - 950$ MHz, appropriate for Band 1  and 2 of the SKA1-MID. We assume that the data over the full band will be sliced into smaller frequency bandwidths of 32 MHz, thus ensuring that each such slice is not affected by light cone effects too severely (i.e., the HI signal can be assumed not to evolve significantly over the 32 MHz band). This allows us to probe the signal at about 13 redshifts, the values of which are taken to be $ 0.5 $, $ 0.6 $, $ 0.7 $, $ 0.8 $, $ 0.9 $, $ 1.0 $, $ 1.1 $, $ 1.3 $, $ 1.5 $, $ 1.7 $, $ 1.9 $, $ 2.2 $ and $ 2.5 $. Note that these are chosen so that the frequency difference between two adjacent redshifts are $\sim 32$ MHz. We assume 1000 hours of observation as the default value across the full band. We also assume that the power spectrum is binned in logarithmic spaced bins in $k$, with $dk / k = 1/5$. The minimum value of $k$ is taken to be 0.005 h $ Mpc^{-1} $ to ensure that the Newtonian perturbation theory is valid and the maximum value of $k$ is taken to be 0.2 h $ Mpc^{-1} $ to ensure that we work in the regime of the linear perturbation theory.

\section{Results}

In this section we present the main results of our work which is to compute the constraints on dark energy models that can be possibly obtained by the upcoming SKA1-mid under ideal conditions. For this we assume two different fiducial models:  first one is the $\Lambda$CDM with $w(z) = -1$ for dark energy EoS; for the second one, we assume that the dark energy EoS is parametrized by CPL parametrization with $w_{0} = - 0.95$ and $ w_{a}= - 0.31$. We call this {\bf fiducial model 2} in our subsequent discussions. This is the best fit CPL model for Planck+BAO+SN+HST without lensing \citep{ade2}.

The 21 cm power spectrum for these fiducial models can be taken to be the proxy for the observational data expected from SKA1-mid. The errors on the observations are estimated using the formalism discussed in the previous section.  So we have two sets of projected data for 21cm observations by SKA1-mid.

Using these two sets of projected data for SKA1-mid, we constrain the three dark energy parametrizations as described in equations (2).

For each model, the parameter constraints are obtained by minimizing the total $ \chi^{2} $ which is defined as

\begin{equation}
\chi^{2} (w_{0},w_{a}) = \sum_{i=1}^{13} \chi^{2}_{i}(w_{0},w_{a}),
\label{eq:chitot}
\end{equation}

\noindent
where summation index $ i $ corresponds to the each redshift bin mentioned in the previous section. At a particular $ i $th redshift bin the $ \chi^{2} $ is defined as

\begin{equation}
\chi^{2}_{i} (w_{0},w_{a}) = \sum_{k} \frac{[P_{21}(k,z_{i},w_{0},w_{a})-P^{fd}_{21}(k,z_{i})]^{2}}{[\delta P^{fd}_{tot} (k,z_{i})]^{2}},
\label{eq:ithchi}
\end{equation}

\noindent
where $ P_{21} $ is calculated using eq. \eqref{eq:P213Davg}. Moreover each redshift bin has all $k$ bins mentioned in the previous section. 

\begin{figure*}[tbp]
\begin{center}
\includegraphics[width=.45\textwidth]{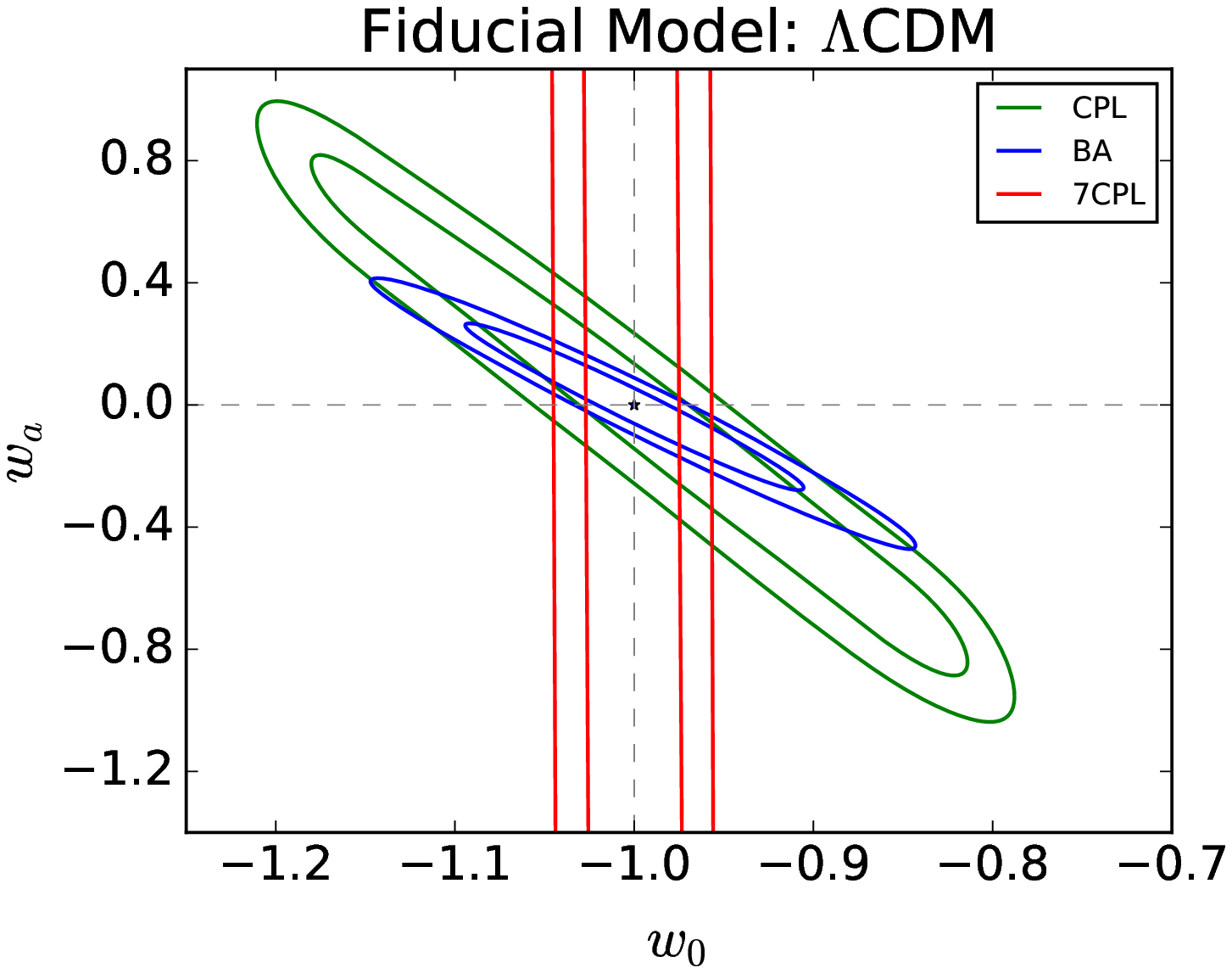}
\includegraphics[width=.45\textwidth]{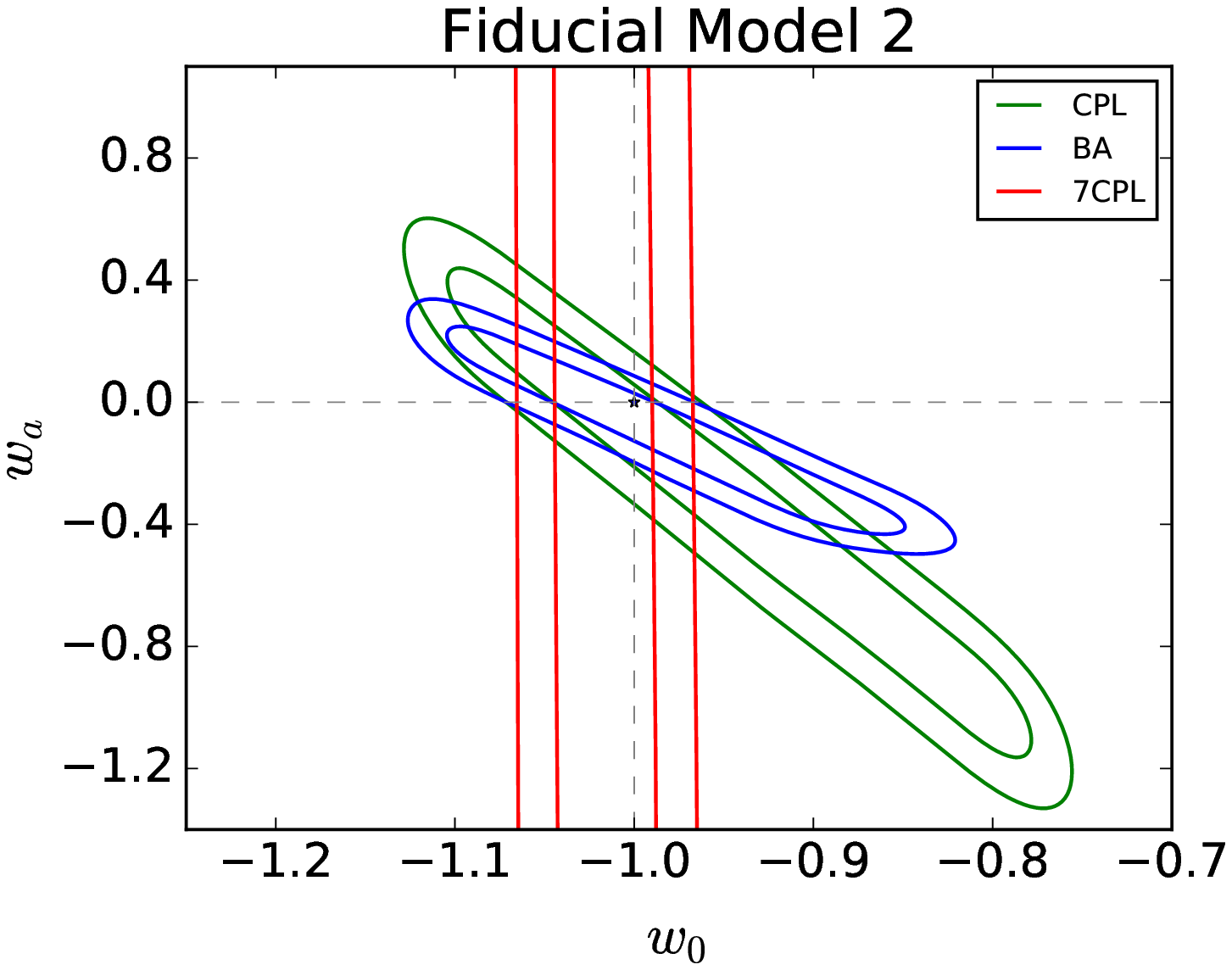}
\caption{\label{fig:cmpcontour} Projected constraints in $ w_{0} $-$ w_{a} $ parameter plane for the three parametrizations. {\it left} for $\Lambda$CDM fiducial model, {\it right} for {\bf fiducial model 2} (see text).}
\end{center}
\end{figure*}

We present our results for two fiducial models described above. The constrained $w_{0}-w_{a}$ parameter space is shown in Figure~\ref{fig:cmpcontour}. There are few interesting observations from these figures:
\begin{itemize}
\item Assuming dark energy is described by CPL parametrization, the SKA-Mid with 1000 hours of observation alone, can give us similar constraint in $w_{0}-w_{a}$ parameter space as obtained by Planck+Bao+Supernova+HST for CPL parametrization (see Figure 4 in \citep{ade2}). Adding other datasets will definitely result much stronger constraint.

\item CPL is not the best parametrization to constraint dark energy evolution. As one can see from Figure~\ref{fig:cmpcontour}, the BA parametrization can constrain the $w_{0}-w_{a}$ parameter space much better than CPL.

\item Although for 7CPL parametrization, the parameter $w_{a}$ is not at all constrained, but 7CPL parametrization can constrain $w_{0}$, the present day dark energy equation of state extremely well, much better than CPL or BA. So if the aim is to measure the present day value of the dark energy equation of state with SKA-Mid, 7CPL is a better choice than CPL or BA. Till date, most of the work aimed to constrain the dark energy evolution, use the CPL parametrization to model the dark energy evolution. This result shows that, in future, one needs to be careful while using CPL parametrization as it may not give the best constraints on the dark energy evolution.

\item For both the fiducial models, the regions ($w_{0} > -1 \& w_{a} >0$) and ($w_{0} < -1 \& w_{a} < 0$) in the parameter space are extremely constrained. In the first region, the dark energy is always non-phantom and in the second region, the dark energy is always phantom and both of these behaviours are severely constrained. On the contrary, the other two regions ($w_{0} > -1 \& w_{a} < 0$) and ($w_{0} < -1 \& w_{a} > 0$) though constrained significantly but still contain larger allowed space compared to the previous ones and in these regions, crossing from non-phantom to phantom and vice versa are allowed. Remember that this phantom crossing is not possible for non interacting single scalar field ( both canonical and non-canonical) dark energy models. To summarise, our study shows that in future, SKA-Mid can severely constrain all single field non-interacting scalar field models which are the most simple and most popular dark energy model beyond $\Lambda$CDM.
\end{itemize}

\begin{figure*}[tbp]
\begin{center}
\includegraphics[width=.45\textwidth]{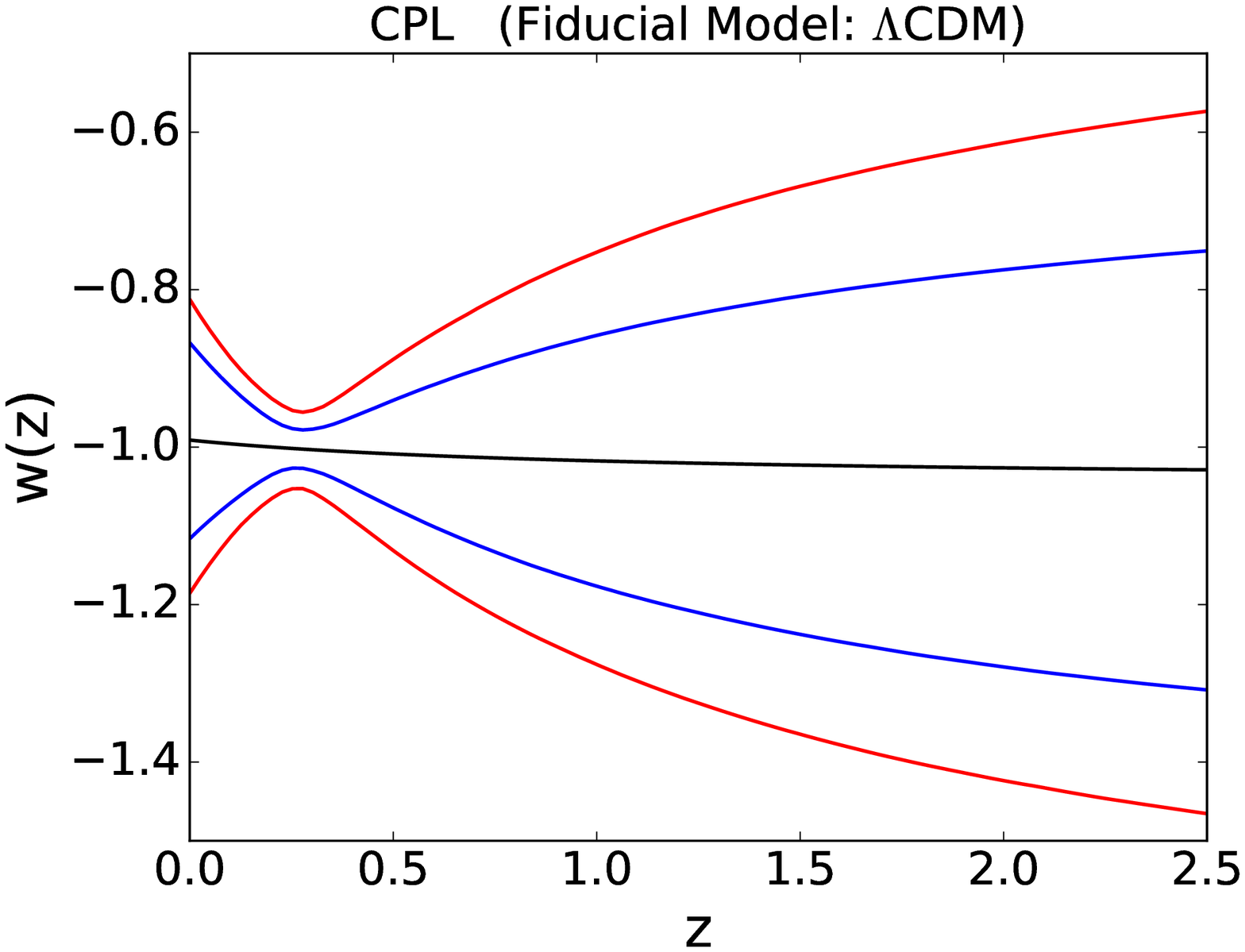}
\includegraphics[width=.45\textwidth]{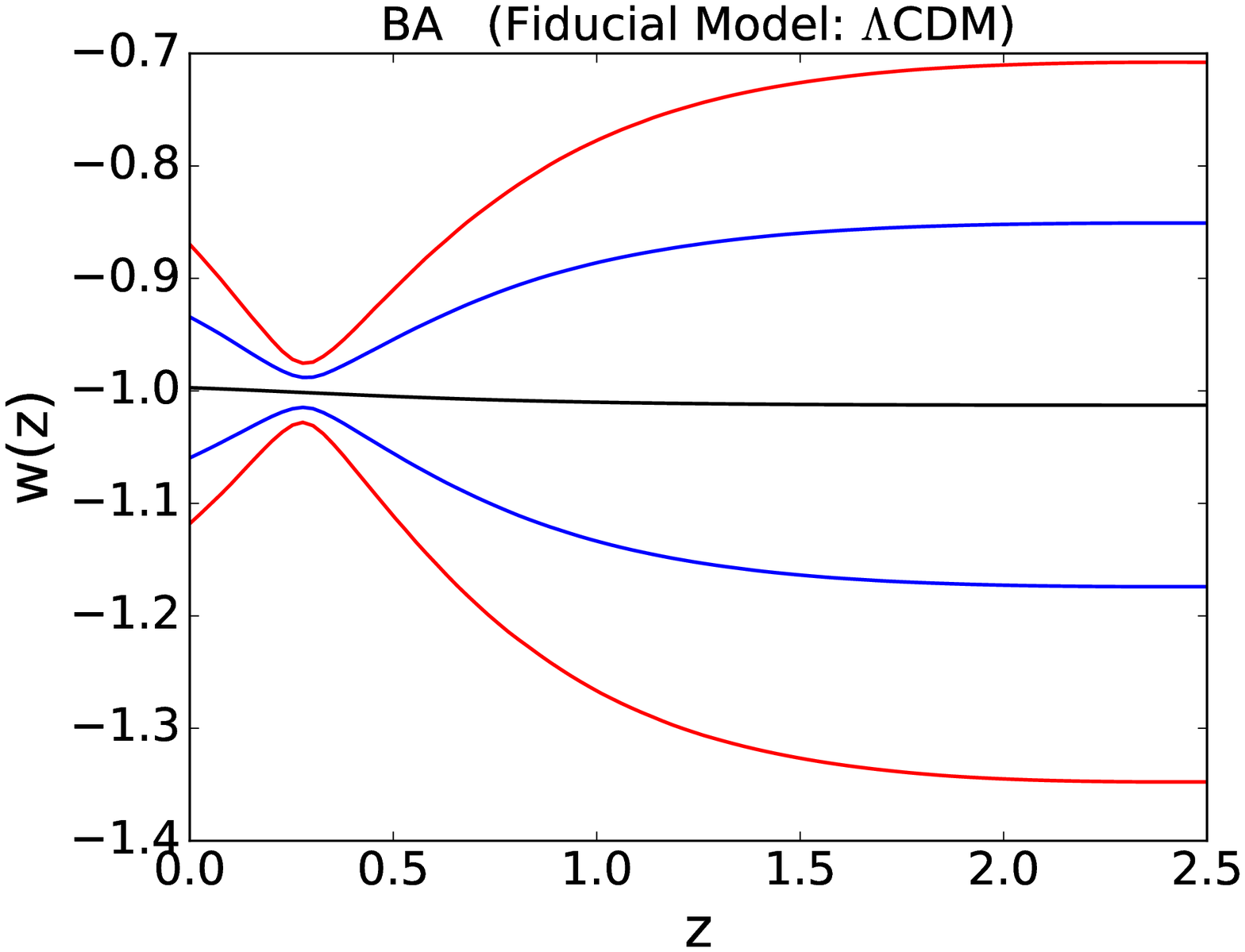}\\
\includegraphics[width=.45\textwidth]{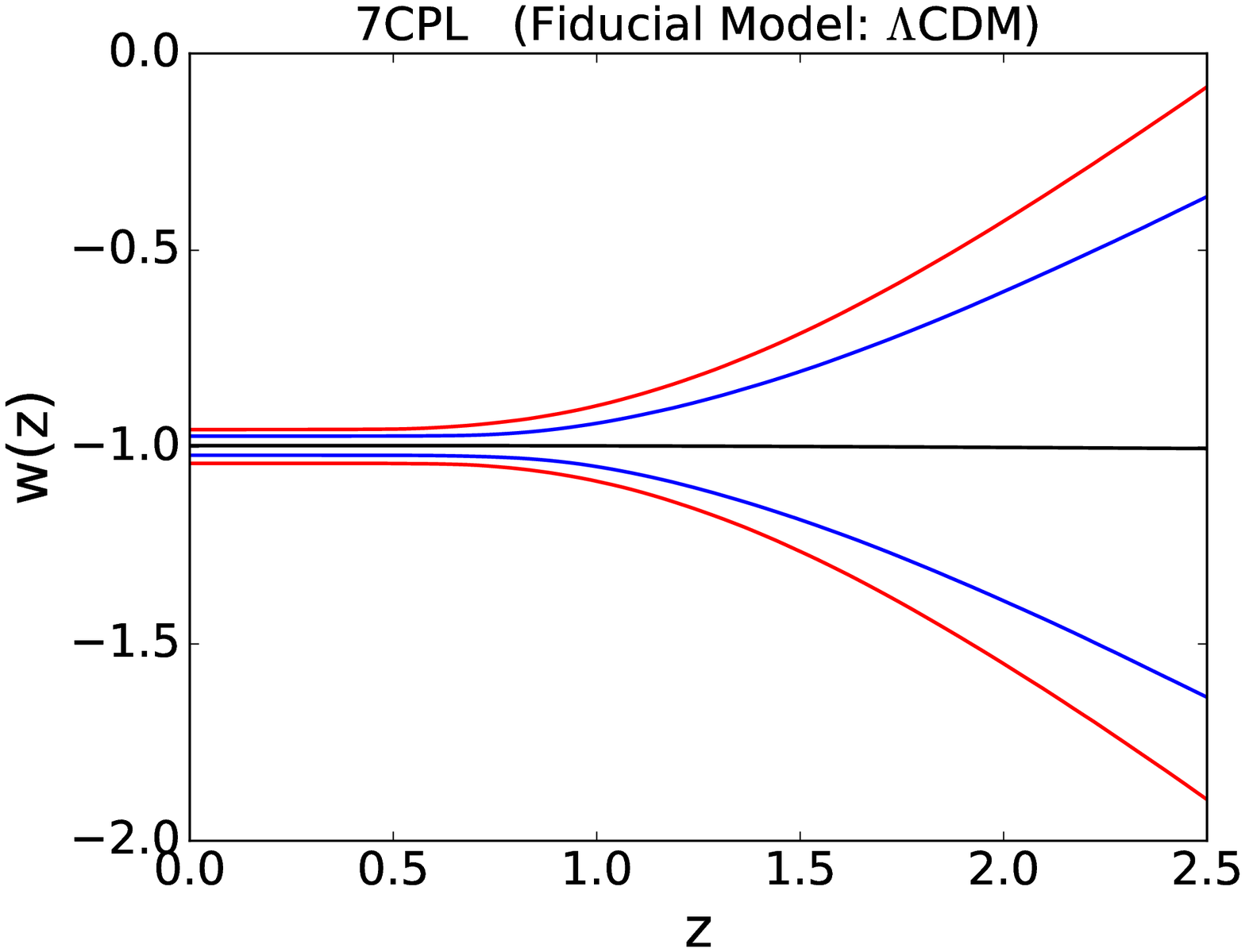}
\caption{\label{fig:derivedLcdm} Reconstructed $ w(z)$ behaviours for different dark energy parametrizations assuming  $\Lambda $CDM as the fiducial model.}
\end{center}
\end{figure*}

\begin{figure}[tbp]
\centering
\includegraphics[width=.45\textwidth]{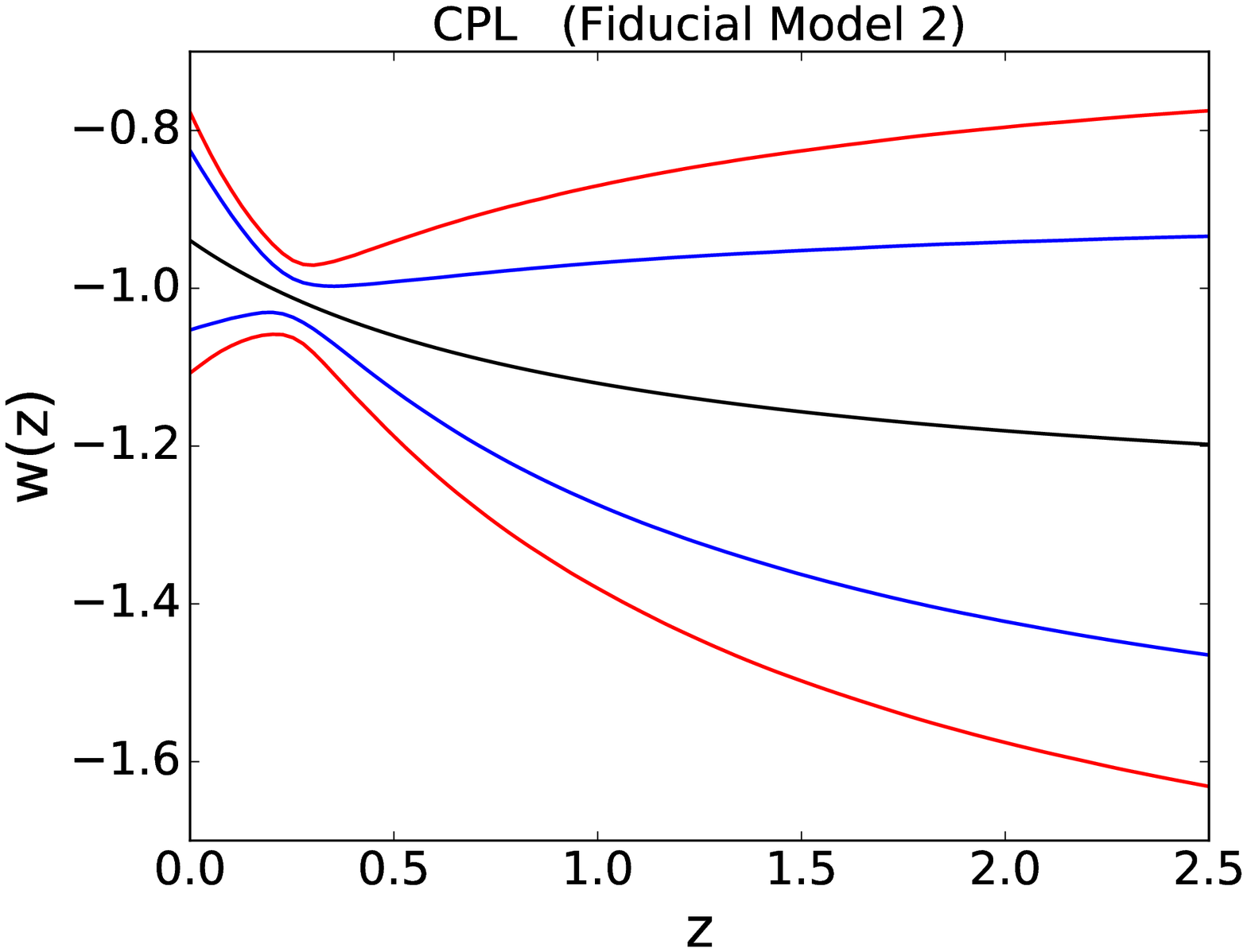}
\includegraphics[width=.45\textwidth]{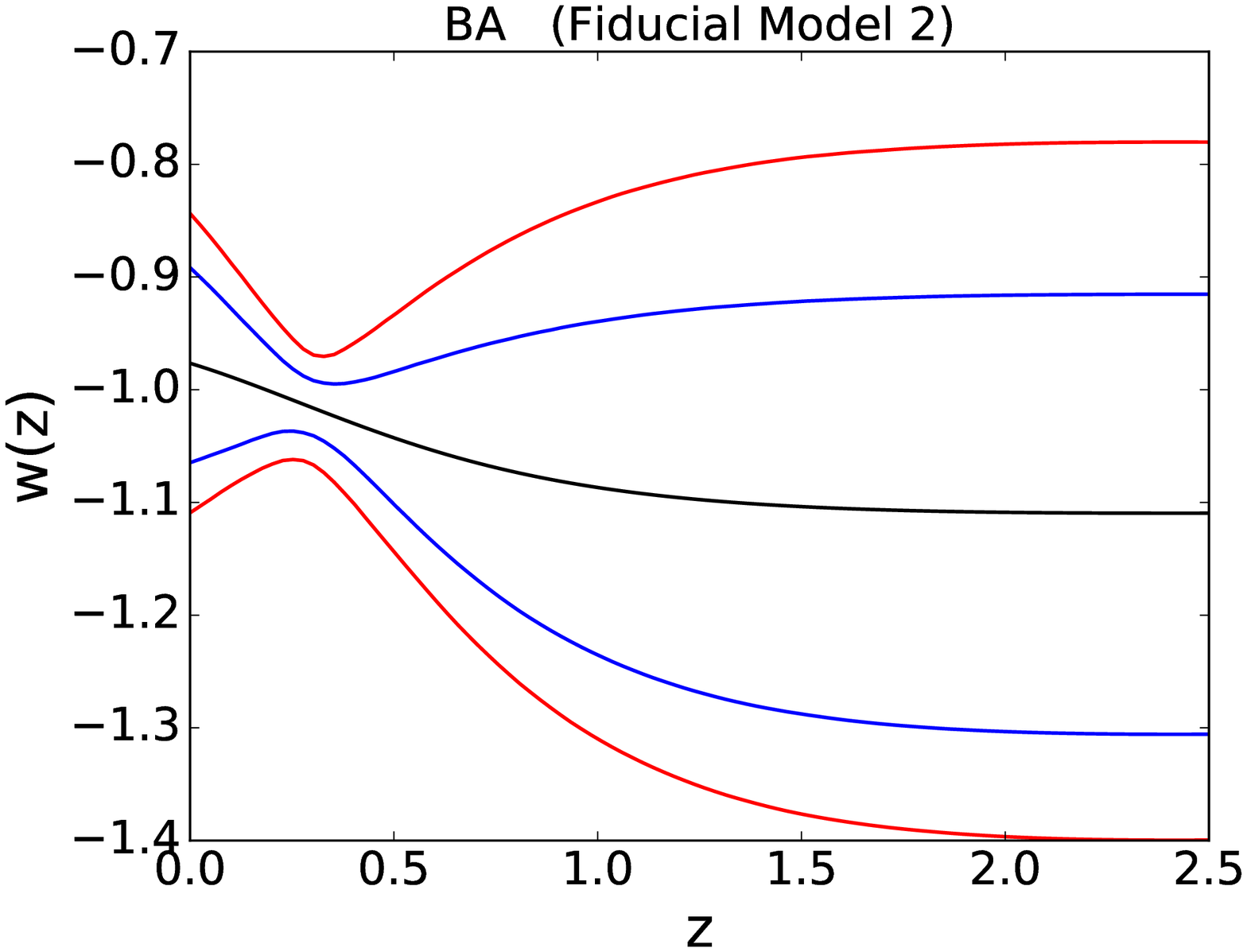}\\
\includegraphics[width=.45\textwidth]{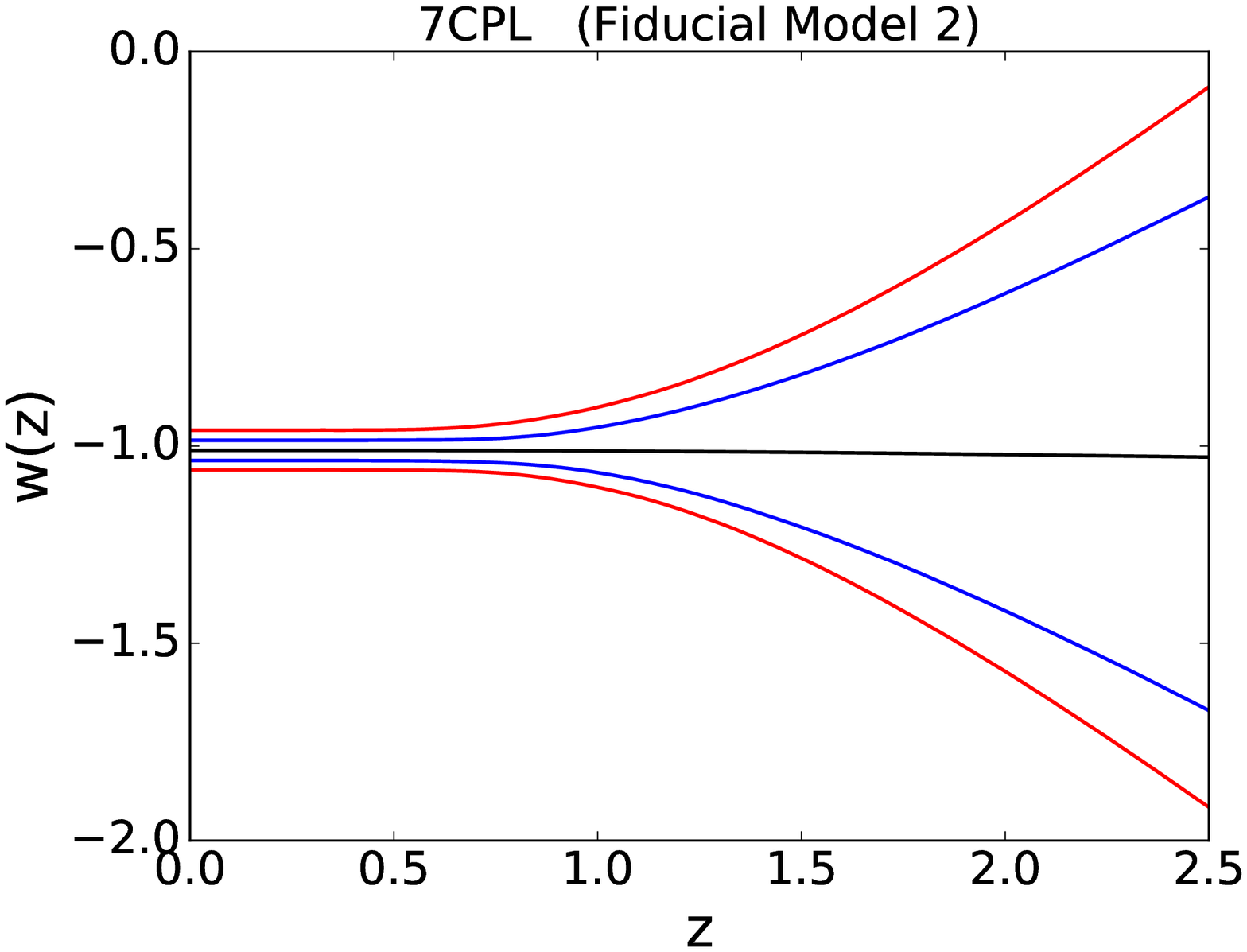}
\caption{\label{fig:derived2} Reconstructed $ w(z)$  behaviours for different dark energy parametrizations for  \textbf{fiducial model 2}.}
\end{figure}

Using the constraints in the $w_{0}-w_{a}$ as shown in Figure~\ref{fig:cmpcontour}, we plot reconstructed $w$ vs. $z$ behaviour for dark energy eos for three different parametrizations used in this study. Figure~\ref{fig:derivedLcdm} is for $\Lambda$CDM fiducial model and Figure~\ref{fig:derived2} is for Fiducial model 2. As we can see, the evolution of the dark energy eos is severely constrained around redshift $z\sim 0.3$ due to the presence of a narrow neck around this redshift. Due to this, both non-phantom ($w>-1$) and phantom behaviour ($w<-1$)  are highly  constrained, whereas a phantom-crossing has slightly better probability. Moreover for 7CPL,  the low redshift revolution of the dark energy eos is severely constrained as compared to CPL and BA. This is consistent with the constrained $w_{0}-w_{a}$ parameter space for 7CPL as shown in Figure~\ref{fig:cmpcontour}.

We also show the $68\%$ and $95\%$ marginalized 1D confidence intervals for the $w_{0}$ and $w_{a}$ parameters for different cases in Tables ~\ref{table:tbl1} and ~\ref{table:tbl2}.

%%%%%%%%%%%%%%%%%%%%%%%%%%%%%%%%%%
%%%%%%%%%%%%%%%%%%%%%%%%%%%%%%%%%%%

\begin{table}[tbp]
\begin{center}
    \begin{tabular}{| l | l | l |}
    \hline
     & $68\%$ marginalized 1D confidence & $95\%$ marginalized 1D confidence \\ \hline
    CPL & $w_{0}^{CPL} = -0.99^{+ 0.16}_{-0.08}$ and $w_{a}^{CPL} = -0.05^{+0.47}_{-0.61}$ & $w_{0}^{CPL} = -0.99^{+0.18}_{-0.19}$ and $w_{a}^{CPL} = -0.05^{+0.89}_{-0.84}$ \\ \hline
    BA & $w_{0}^{BA} = -1.0^{+0.06}_{-0.06}$ and $w_{a}^{BA} = -0.01^{+0.19}_{-0.17}$ & $w_{0}^{BA} = -1.0^{+0.12}_{-0.12}$ and $w_{a}^{BA} = -0.01^{+0.34}_{-0.37}$ \\ \hline
    7CPL & $w_{0}^{7CPL} = -1.0^{+0.03}_{-0.02}$ and $w_{a}^{7CPL} = NIL$ & $w_{0}^{7CPL} = -1.0^{+0.04}_{-0.05}$ and $w_{a}^{7CPL} = NIL$ \\ 
    \hline
    \end{tabular}
\end{center}
\caption{Fiducial Model: $\Lambda$CDM} % title of Table
\label{table:tbl1}
\end{table}

\begin{table}[tbp]
\begin{center}
    \begin{tabular}{| l | l | l |}
    \hline
     & $68\%$ marginalized 1D confidence & $95\%$ marginalized 1D confidence \\ \hline
    CPL & $w_{0}^{CPL} = -0.948^{+0.097}_{-0.097}$ and $w_{a}^{CPL} = -0.311^{+0.46}_{-0.36}$ & $w_{0}^{CPL} = -0.948^{+0.16}_{-0.16}$ and $w_{a}^{CPL} = -0.311^{+0.80}_{-0.81}$ \\ \hline
    BA & $w_{0}^{BA} = -0.977^{+0.092}_{-0.076}$ and $w_{a}^{BA} = -0.11^{+0.17}_{-0.28}$ & $w_{0}^{BA} = -0.977^{+0.13}_{-0.13}$ and $w_{a}^{BA} = -0.11^{+0.35}_{-0.34}$ \\ \hline
    7CPL & $w_{0}^{7CPL} = -1.011^{+0.025}_{-0.025}$ and $w_{a}^{7CPL} = NIL$ & $w_{0}^{7CPL} = -1.011^{+0.050}_{-0.049}$ and $w_{a}^{7CPL} = NIL$ \\ 
    \hline
    \end{tabular}
\end{center}
\caption{Fiducial Model: \textbf{fiducial model 2}.} % title of Table
\label{table:tbl2}
\end{table}

%%%%%%%%%%%%%%%%%%%%%%%%%%%%%%%%%%
%%%%%%%%%%%%%%%%%%%%%%%%%%%%%%%%%%

%%%%%%%%%%%%%%%%%%%%%%%%%%%%%%%%%%
%%%%%%%%%%%%%%%%%%%%%%%%%%%%%%%%%%

\section{Conclusions}

In this paper, we study the projected constrained on dark energy evolution using post-reionization 21cm power spectra from the neutral hydrogen in galaxies as measured by SKA1-mid. To model the dark energy evolution beyond $\Lambda$CDM, we use three different parametrizations for dark energy equation of state including the widely used CPL parametrization. To generate the simulated data, we assume two fiducial models: the concordance $\Lambda$CDM model and the best fit CPL model for Planck+BAO+SN+HST data as obtained in \citep{ade2}. To calculate the projected errors ( both system noise and cosmic variance), we use SKA1-mid specifications with 200 antannae each with 15 meter diameter. The total observation time is assumed to be 1000 hours. We assume the Band-1 and Band-2 frequency range for SKA1-mid with total 13 redshift bins.

The projected constraints on the dark energy equation of state have several interesting features. We show that SKA1-mid alone can constrain dark energy evolution at the same level of joint Planck+SN+BAO+HST constraints at present. We also show that the single non-interacting scalar field models (both canonical and non-canonical) can be severely constrained by SKA1-mid type observations. There is a larger parameter space for models with phantom crossing. Moreover if we aim to constrain the present value of the dark energy equation of state, 7CPL parametrization is much better than CPL and BA parametrization. Also CPL parametrization which is the widely used parametrization to model dark energy evolution, may not be suitable to constrain  dark energy evolution. BA parametrization can deliver much better constraints for dark energy evolution.

To conclude, SKA1-mid has the potential to shed new lights on dark energy evolution. Combining with other observational results from CMB, BAO, SNIa etc, it will hopefully tell us whether the dark energy is cosmological constant or not, whether it is phantom or non-phantom and whether there is any phantom crossing in the dark energy evolution. But we need to be also careful about how we model the dark energy evolution, because as shown in this study, not all parametrizations can achieve the required accuracy to constrain the dark energy evolution.

\section{Acknowledgements}
B.R.D. thanks CSIR, Govt. of India for financial support through SRF scheme (No:09/466(0157)/2012-EMR-I). A.A.S thanks IUCAA Pune, India where part of the work has been done, for his visit through Associateship program. The authors also thank Dr. Kanan Dutta for discussions.

\end{document}